\documentclass[english,reprint,twocolumn,amsmath,amssymb,prd,nofootinbib,superscriptaddress]{revtex4-2}
\usepackage[T1]{fontenc}
\usepackage[latin9]{inputenc}
\usepackage{geometry}
\usepackage{graphicx}
\usepackage{bm}
\usepackage{hyperref}
\hypersetup{colorlinks, citecolor=Blue, linkcolor=Blue, urlcolor=Blue}
\usepackage[normalem]{ulem}
\usepackage[usenames,dvipsnames]{color}
\usepackage[sort&compress]{natbib}
\usepackage{textcomp}
\usepackage{babel}
\usepackage{siunitx}
\usepackage{amsmath,amssymb}
\usepackage{lipsum}
\newcommand{\orcid}[1]{\href{https://orcid.org/#1}{\textcolor{black}{\,\includegraphics[height=2ex]{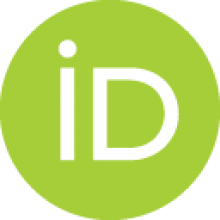}}}}

\begin{document}
\setlength{\parskip}{0pt}
\title{Ultra-Heavy Dark Matter Search with Electron Microscopy of Geological Quartz}

\author{Reza~Ebadi\orcid{0000-0003-1325-6876}\,$^\dagger$}
\email{ebadi@umd.edu}
\affiliation{Department of Physics, University of Maryland, College Park, Maryland 20742, USA}
\affiliation{Quantum Technology Center, University of Maryland, College Park, Maryland 20742, USA}
\noaffiliation
\author{Anubhav~Mathur\orcid{0000-0003-0973-1793}}
\thanks{These authors contributed equally to this work.}
\affiliation{Department of Physics and Astronomy, Johns Hopkins University, 3400 N. Charles St., Baltimore, Maryland 21218, USA}
\author{Erwin~H.~Tanin\orcid{0000-0002-3204-0969}}
\thanks{These authors contributed equally to this work.}
\affiliation{Department of Physics and Astronomy, Johns Hopkins University, 3400 N. Charles St., Baltimore, Maryland 21218, USA}
\author{Nicholas~D.~Tailby}
\affiliation{Department of Earth and Planetary Sciences, American Museum of Natural History, New York, New York 10024, USA}
\author{Mason~C.~Marshall\orcid{https://orcid.org/0000-0002-7089-2548}}
\affiliation{Quantum Technology Center, University of Maryland, College Park, Maryland 20742, USA}
\affiliation{Department of Electrical and Computer Engineering, University of Maryland, College Park, Maryland 20742, USA}
\affiliation{Harvard-Smithsonian Center for Astrophysics, Cambridge, Massachusetts 02138, USA}
\author{Aakash~Ravi\orcid{0000-0003-1312-8161}}
\affiliation{Quantum Technology Center, University of Maryland, College Park, Maryland 20742, USA}
\author{Raisa~Trubko}
\affiliation{Department of Physics, Worcester Polytechnic Institute, Worcester, Massachusetts 01609 USA}
\affiliation{Department of Physics, Harvard University, Cambridge, Massachusetts 02138, USA}
\affiliation{Department of Earth and Planetary Sciences, Harvard University, Cambridge, Massachusetts 02138, USA}
\author{Roger~R.~Fu}
\affiliation{Department of Earth and Planetary Sciences, Harvard University, Cambridge, Massachusetts 02138, USA}
\author{David~F.~Phillips\orcid{0000-0001-5132-1339}}
\email{dphillips@cfa.harvard.edu}
\affiliation{Harvard-Smithsonian Center for Astrophysics, Cambridge, Massachusetts 02138, USA}
\author{Surjeet~Rajendran\orcid{0000-0001-9915-3573}}
\email{srajend4@jhu.edu}
\affiliation{Department of Physics and Astronomy, Johns Hopkins University, 3400 N. Charles St., Baltimore, Maryland 21218, USA}
\author{Ronald~L.~Walsworth\orcid{https://orcid.org/0000-0003-0311-4751}}
\email{walsworth@umd.edu}
\affiliation{Department of Physics, University of Maryland, College Park, Maryland 20742, USA}
\affiliation{Quantum Technology Center, University of Maryland, College Park, Maryland 20742, USA}
\affiliation{Department of Electrical and Computer Engineering, University of Maryland, College Park, Maryland 20742, USA}
\date{\today}
\begin{abstract}
Self-interactions within the dark sector could clump dark matter into heavy composite states with low number density, leading to a highly suppressed event rate in existing direct detection experiments. However, the large interaction cross section between such ultra-heavy dark matter (UHDM) and standard model matter results in a distinctive and compelling signature: long, straight damage tracks as they pass through and scatter with matter. In this work, we propose using geologically old quartz samples as large-exposure detectors for UHDM. We describe a high-resolution readout method based on electron microscopy, characterize the most favorable geological samples for this approach, and study its reach in a simple model of the dark sector. The advantage of this search strategy is two-fold: the age of geological quartz compensates for the low number density of UHDMs, and the distinct geometry of the damage track serves as a high-fidelity background rejection tool.
\end{abstract}
\maketitle

\def\beq{\begin{equation}}
\def\eeq{\end{equation}}

\def\bea{\begin{eqnarray}}
\def\eea{\end{eqnarray}}

\def\Mp{M_{\rm pl}}

\def\rv{{\bf r}}
\def\Rv{{\bf R}}
\def\kv{{\bf k}}
\def\kvp{{\bf k}^{\hskip 1pt \prime}}
\def\qv{{\bf q}}
\def\pv{{\bf p}}
\def\vv{{\bf v}}
\def\xv{{\bf x}}
\def\xvp{{\bf x}^{\hskip 1pt \prime}}
\def\yv{{\bf y}}
\def\Pv{{\vec P}}
\def\uv{{\vec u}}

\def\DC{{\cal D}}
\def\HC{{\cal H}}
\def\LC{{\cal L}}
\def\OC{{\cal O}}


\def\eV{{\rm eV}}
\def\keV{{\rm keV}}
\def\MeV{{\rm MeV}}
\def\GeV{{\rm GeV}}
\def\TeV{{\rm TeV}}

\def\angs{\SI{}{\angstrom}}
\def\nm{{\rm nm}}
\def\micron{\SI{}{\micro\meter}}
\def\micro{\SI{}{\micro}}
\def\mm{{\rm mm}}
\def\cm{{\rm cm}}
\def\m{{\rm m}}
\def\km{{\rm km}}
\def\gm{{\rm gm}}
\def\s{{\rm s}}

\def\C{{\rm C}}

\def\MJ{{\rm MJ}}
\def\pJ{{\rm pJ}}

\def\d{{\rm d}}
\def\a{{\rm a}}
\def\b{{\rm b}}

\def\melt{{\rm melt}}
\def\WIMP{{\rm WIMP}}
\def\blob{{\rm blob}}
\def\si{{\rm SI}}
\def\SD{{\rm SD}}
\def\DM{{\rm DM}}
\def\N{{\rm n}}
\def\p{{\rm p}}
\def\g{{\rm g}}
\def\yr{{\rm yr}}
\def\Gyr{{\rm Gyr}}
\def\month{{\rm month}}

\def\psibar{\bar{\psi}}
\def\chibar{\bar{\chi}}

\section{Introduction}

A major outstanding puzzle in modern physics is the nature of dark matter (DM) \cite{pdg2018}. Despite the ever-improving sensitivities of direct detection experiments, the simplest DM candidates have not been observed, motivating searches for a wider range of possible dark sectors. Moreover, challenges that simple cold DM candidates face on sub-galactic scales \cite{Kuhlen:2012ft,Bullock:2017xww} might be relieved with more complex dark sectors. For instance, self-interacting DM has been investigated to address small-scale challenges such as the core-cusp problem \cite{Spergel:1999mh,Loeb:2010gj}.

Nearly all current DM detection strategies, ranging from direct-detection efforts in the laboratory to indirect signals from DM annihilation (or decay), are based on the assumption that the DM is distributed around the universe as a gas of free particles with a large number density. This picture naturally emerges if self-interactions within the dark sector are weak, but is not strictly prescribed by existing observational constraints. The most stringent limits arise from observations of the Bullet Cluster, which restrict the self-interaction cross section per unit mass to be $\sigma_{\chi\chi}/m_\DM \lesssim 1 ~\cm^2/\g$ \cite{Spergel:1999mh}. It is apparent that the limit on $\sigma_{\chi\chi}$ itself is significantly weakened if DM is clustered into composite states with large masses $m_\DM$.

If the dark sector has strong self-interactions, it would undergo a nucleosynthesis process in the early universe much like the nuclei of the standard model (SM), whereby individual particles coalesce to form large composite states \cite{Hardy:2014mqa,Wise:2014ola}. SM nucleosynthesis suffers from a number of theoretical accidents (such as the deuterium bottleneck) that render certain light elements unstable and thereby inhibit the production pathway of ultra-heavy elements. Still, the SM manages to produce large composite systems \cite{ghiorso1955new}. It is thus not surprising that a completely unconstrained dark sector could also produce large composite objects. We will refer to these composite states, which can easily be much heavier than $10^{24}~\GeV$, as ultra-heavy dark matter \cite{Gresham:2017cvl,Grabowska:2018lnd}.

Direct searches for canonical DM in the form of a gas of small particles leverage the large influx of these particles in a detector to compensate for their small cross section with SM particles. For example, about $10^{16}$ weakly interacting massive particles (WIMPs) of mass 100 GeV would pass through a $\m^3$ detector in a year, allowing for the direct detection of WIMP-nucleon scattering cross sections as low as $10^{-45}\text{ cm}^2$ \cite{Roszkowski_2018}. UHDMs, on the other hand, would arrive with significantly lower flux and  require a different detection strategy. Thus, an experimental strategy for UHDM detection should leverage generic signatures of large composite objects instead of focusing on the specifics of any one composite DM model, as the rich dynamics of an interacting dark sector can produce a plethora of models. Accordingly, we focus here on the fact that the many constituent particles of a UHDM will enhance its cross section with SM matter such that each rare UHDM incidence could result in a spectacular (and possibly lethal \cite{Sidhu:2019oii}) event with a distinct signature imitated by no SM effect. The detector sensitivity is therefore less important in such scenarios and can be traded for greater exposure that maximizes the probability of a rare UHDM transit. In this paper, our primary focus is to establish the detectability of such a signal. While we provide an example of a DM model that yields such a signal, it is straightforward to construct other examples of ultra-heavy particles that can cause similar damage (e.g., Q-balls \cite{Kusenko:1997si}, charged primordial black hole relics \cite{Lehmann:2019zgt}). 

Damage tracks left by particles passing through the Earth over geological times could be recorded by ancient rock samples buried underground. For this reason, such geological samples have been proposed and used as "natural particle detectors" in the past, including for magnetic monopoles \cite{Fleischer1969,Eberhard1971,Kovalik1986,Jeon:1995rf,fleischer1969jan,fleischer1969aug,price1984,price1986}, macroscopically large DM \cite{Sidhu:2019qoa}, WIMPs \cite{Snowden-Ifft1995,Collar:1994mj,engel1995,SnowdenIfft:1997hd,Baum:2018tfw,Drukier:2018pdy,Edwards:2018hcf}, and neutrinos \cite{Baum:2019fqm,Jordan:2020gxx}. The geological exposure times of ancient rock detectors range from the present back to the time they were last heated naturally to the point of annealing, which can be up to $\sim 10^9$ years and is thus much longer than typical DM direct detection laboratory experiments (by factors of up to $10^9$). This advantage makes geological DM detectors ideal probes of sparser, higher-mass composite DM. As we discuss in Sec. \ref{sec:feasibility}, searches with $10 ~\m^2$ of billion-year old rock would probe DM masses up to $m_\DM \sim 10^{28}~\GeV$. A major challenge for such a detection strategy is the ability to efficiently identify DM signatures in a large volume of rock and distinguish them from geological, radioactive, and cosmic ray backgrounds. Such discrimination is significantly simpler in searches for UHDMs, since the extremely long and continuous cylindrical damage patterns they generically leave are qualitatively different from the sporadic defects due to expected backgrounds.

Here, we assess the use of geologically old quartz samples as solid-state particle detectors to search for damage tracks left by UHDMs. Quartz, a crystalline polymorph of silica SiO$_2$, is one of the most abundant and well-studied minerals in the lithosphere \cite{gotze2012application}. Defects and damage tracks can be resolved down to the micron scale with SEM-CL: a scanning electron microscope (SEM) combined with a cathodoluminescence (CL) detector. Imaging provided by the SEM is supplemented with spectral information from CL, which reveals the nature of trace elements and point defects in the quartz \cite{stevens2009cathodoluminescence}.  This modality has already proved successful at providing answers to key geological questions \cite{vasyukova2013,macrae2013hyperspectral,leeman2012study,spear2009cathodoluminescence,ackerson2018low,hamers2017scanning,ackerson2015trace}. The technical advantages of SEM-CL mapping, as well as the considerable literature on its application in quartz, make it an appropriate choice for our readout method.

Note that a similar search for long damage tracks was performed by Price and Salamon \cite{price1986} in ancient mica crystals with null results. While they used this result to constrain the abundance of magnetic monopoles, the experiment is also sensitive to UHDMs with masses $m_{\rm DM}\lesssim 10^{26}\,\GeV$ \cite{Jacobs_2015,Bhoonah:2020dzs}. Mica is a crystalline mineral with almost perfect basal cleavage, offering efficient sample preparation. However, the readout method used in their experiment requires acid etching prior to microscopy in order to enlarge damage tracks and render them visible in an optical microscope. Such an etching process also enlarges background signals in the form of naturally-occurring lattice defects. Hence, the success of the experiment hinges on the low level of background in the samples and its scalability is limited by the availability of sufficiently pristine mica crystals. By contrast, the lack of etching in our proposed SEM-CL readout and the more readily available quartz mineral that meets our requirements allow us to scan over larger sample areas and extend the search pioneered by Price and Salamon to higher DM masses.

The rest of the discussion is organized as follows. In Sec.~\ref{sec:feasibility} we present an experimental realization of the UHDM detection method sketched above, identify optimal samples, and assess its model-independent sensitivity. We then demonstrate its ability to probe a simple composite UHDM model, taking into account various existing constraints, in Sec.~\ref{sec:model}. Finally, we conclude in Sec.~\ref{sec:conclusion}.

\section{Detection Feasibility}
\label{sec:feasibility}
In this section, we investigate experimental issues for UHDM detection with quartz. Based on the considerations below, we establish the criteria for UHDM signatures to be robustly detectable with microscopy of about $\micron$-resolution. We discuss the discovery reach of this approach in Sec. \ref{sec:sensitivity} (see equations \eqref{eqn:dragforcebound}-\eqref{eqn:meltingenergypernucleus}).

\begin{figure}[htbp!]
    \centering
    \includegraphics[width=1\columnwidth]{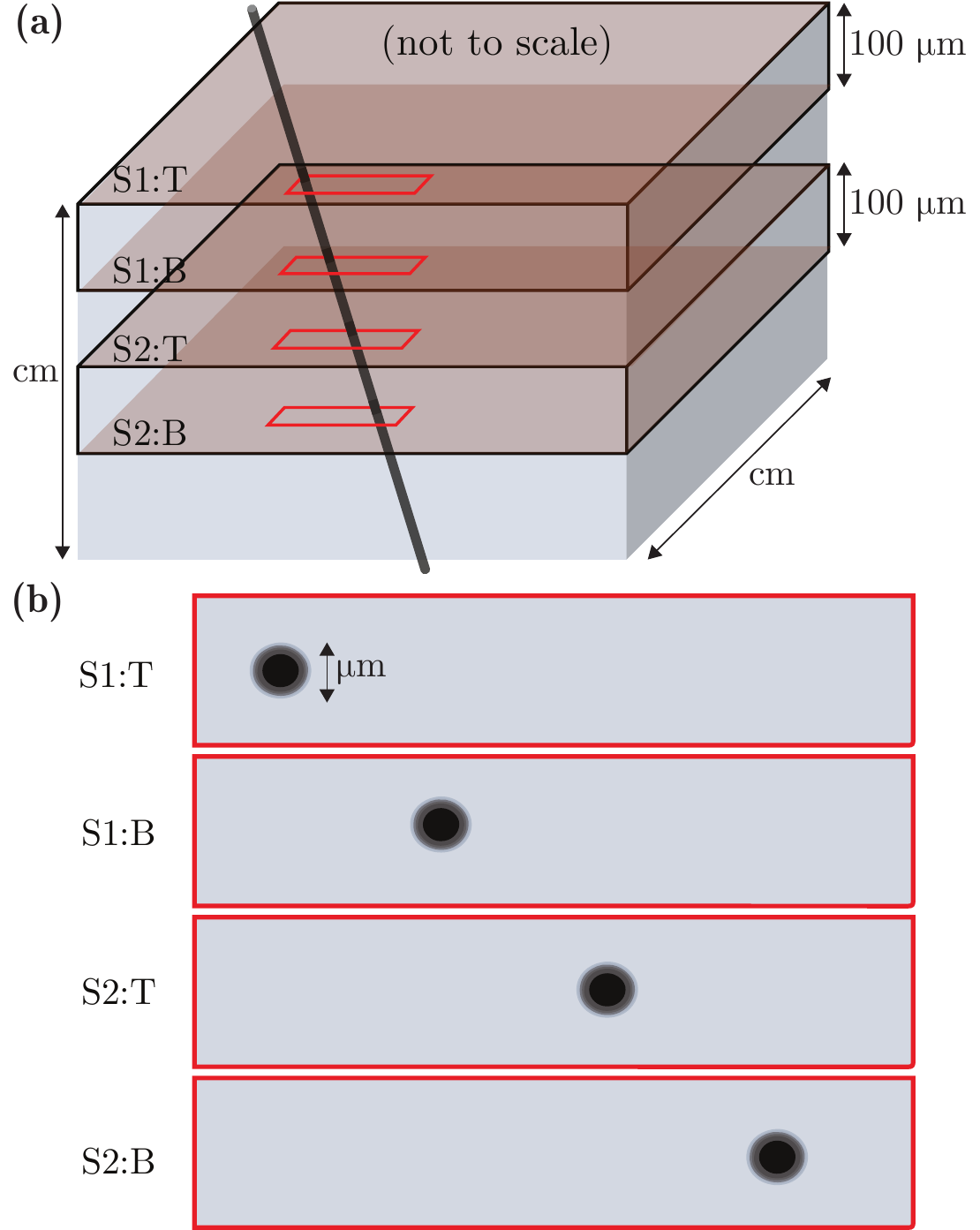}
    \caption{{Schematic of the proposed readout method.} {\bf (a)} A quartz sample of size $\sim\cm^3$. The black straight line illustrates a damage track as a result of an ultra-heavy composite dark matter (UHDM) particle passing through the sample. The sample is sectioned into multiple sections of thickness $\sim100~\micron$. We show several sections where the top and bottom surfaces are highlighted, which would be scanned using SEM-CL.  {\bf (b)} Correlated damage spots of micron-scale diameter over a macroscopic (mm-scale or longer) distance, between sections is the unique signature of the ultra-heavy DM particle interaction with quartz. Note that the probability of background features coincidently aligning reduces exponentially with the number of correlated layers. For a realistic feature density of $1000/\cm^2$, simulations show that correlations of 4 layers efficiently rejects false positive signals.}
    \label{fig:readout}
\end{figure}

\subsection{Damage Tracks}
Solid-state systems \cite{fleischer1964,fleischer1965,lannunziata2012} have been used as particle track detectors with applications ranging from nuclear science to geophysics, as such tracks yield information about the history of the sample and properties of the impinging particles \cite{fleischer1965a}. For example, DM detection using crystal damage tracks has been proposed as a directional signal in semiconductors such as diamond, where a WIMP scattering event would give rise to damage tracks of $\sim ~30-100 ~\nm$ in length; the directional information in these tracks could enable such a detector to probe below the "neutrino floor" \cite{Rajendran:2017ynw,Marshall:2020azl}. Paleodetection, which looks for damage tracks of a similar size in ancient rock samples, has also been investigated for WIMP detection \cite{Baum:2018tfw,Drukier:2018pdy,Edwards:2018hcf}.

We propose using ancient quartz as a detector for ultra-heavy composite dark matter (UHDMs). As we discuss in Sec. \ref{sec:sensitivity}, each UHDM could deposit enough energy to locally melt nearby quartz along its trajectory. Since quartz nucleation under ambient conditions is a very slow process \cite{buckley2018nucleation}, the melted region would solidify into amorphous silica without recrystallizing. Detecting such amorphous micro-regions within quartz samples is feasible with SEM-CL, where defects in the tetrahedrally coordinated SiO$_2$ microstructure contribute to CL emission \cite{stevens2013cathodoluminescence}. Even in the absence of melting, the same SEM-CL method would in principle be sensitive to linear tracks of lattice distortions left by UHDMs. However, quantitatively characterizing the sensitivity of this method to such tracks requires further study of the backgrounds in natural quartz, which we leave for future work. Thus, in what follows we focus on detecting UHDMs that can cause melting.

\subsection{Quartz Samples and Backgrounds}
\label{sec:background}

The signature of the proposed UHDM detection method is a long damage track, of micron-scale cross section, extending through the entire length of the quartz sample (see Figure \ref{fig:readout}a). This signature has the distinct advantage that no known mechanism produces such a track, allowing for strong geometric rejection of background signals. A variety of effects may induce localized disruption of the crystal lattice on the micron scale, such as extended growth defects or radioactive decays, but these localized features cannot pass through the entire macroscopic crystal. And although particles with low interaction probability (e.g., neutrinos or relativistic cosmic rays) can pass through an entire sample, their low nuclear cross sections yield dispersed individual damage events rather than a continuous, micron-scale-diameter track. The density of atmospheric neutrino-induced damage tracks falls off exponentially with the track length, with essentially zero density of tracks longer than a few mm \cite{Jordan:2020gxx}. Note that the imparted energy per unit length along such tracks is much lower than the robust detection threshold; however, if detected, they can be rejected as a UHDM signal due to the lack of correlation throughout the full $\sim$ cm extent of the sample.

Fractures induced by historical geological stresses could similarly pass continuously through an entire sample, but would in general be two- or three- rather than one-dimensional, and would not leave behind amorphous quartz within the damage track. Geologically-induced single line defects might also be present in quartz samples, however, at smaller scales, only observable by transmission electron microscopy (TEM) \cite{leroux1994tem} or atomic force microscopy (AFM). So these features are not relevant backgrounds for the proposed readout.

\begin{figure*}
\includegraphics[width=2\columnwidth]{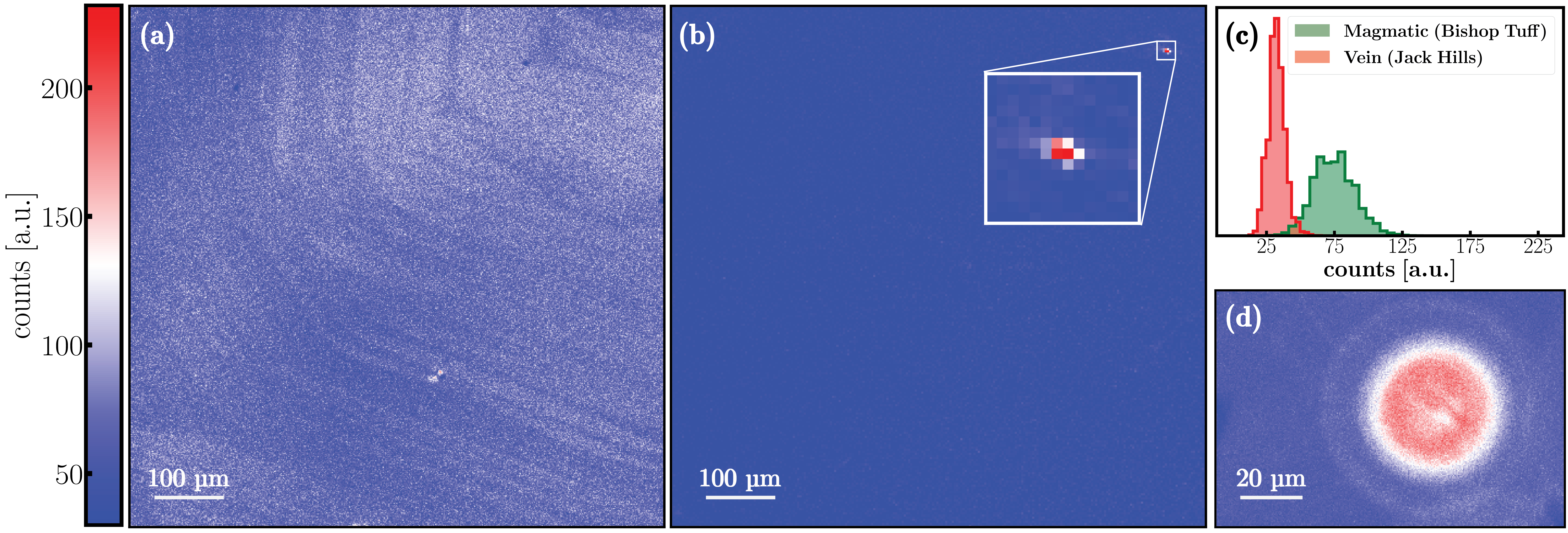}
\caption{\label{fig:CLimage}Example quartz sample characterization. SEM-CL images of two samples, {\bf (a)} magmatic quartz from Bishop Tuff with Ti concentration $51 \pm 6$ ppm, and {\bf (b)} vein quartz from Jack Hills with Ti concentration $5.2 \pm 6.5$ ppm, measured on a mass spectrometer. The scan rate is $20~\s/\mm^2$ with $1.5~\micron$ resolution for magmatic quartz and $5~\s/\mm^2$ with $3~\micron$ resolution for vein quartz (we forecast the full-scale UHDM experiment time and resources using these values). In (b) we identify a few high-count pixels in the vein quartz image, which demonstrates the possibility of high-resolution detection of concentrated CL emission. The inset shows a zoomed-in image of the region of interest with high-count pixels. These pixels could be a melting track intersection, which needs to be investigated by correlating multiple sections as described in the text. {\bf (c)} Normalized histogram of the pixel counts in arbitrary units for each of the two sample SEM-CL images. Vein quartz shows a lower CL noise level as well as smaller variation, making it a suitable target for our detection proposal. {\bf (d)} SEM-CL signal from a uranium halo (measured in a different quartz sample from those shown in (a) and (b)).  Microscopic uranium inclusions have decayed over time; the fission products from these inclusions create crystal lattice damage, which emits cathodoluminescence (CL) upon excitation by the SEM. The CL signal from an ultra-heavy composite dark matter (UHDM) particle track would also result from crystal lattice defects at and around the track of melted quartz.  Any such uranium halos in a UHDM search would be disqualified as potential damage tracks by lack of correlated damage in other slices of the sample.}
\end{figure*}

To take advantage of the distinctive extended geometry of a UHDM signal, we propose a multi-phase scanning readout, where we search for correlated feature positions at multiple depths in the sample (see Figure \ref{fig:readout}). This allows us to reject backgrounds, which have an exponentially suppressed likelihood of lying along a single line  (as shown by simulations discussed in Figure \ref{fig:readout}). An expected background signal, given the imaging resolution of our method, is from  the presence of radioactive isotopes such as uranium, which lead to fission tracks and alpha recoil damage within the crystal lattice. These processes leave behind halos of size $\sim~10~ \micron$ \cite{Bower2016}, which are readily detectable using our proposed SEM-CL protocol (see Figure \ref{fig:CLimage}d). In a single two-dimensional SEM-CL scan, these could mimic a UHDM damage track, but would be disqualified as UHDM signals by lack of correlated damage in subsequent slices. The presence of some radioactivity-induced features is potentially beneficial to our analysis, as their preservation would indicate that recent annealing events would not have removed older, UHDM-induced features. As such, if the fission track age can be determined from the host quartz, the absence of UHDM-induced features implies the lack of UHDM interaction events since the time of occurrence of the fission track.

Quartz samples with low impurity levels are essential for reducing background levels. High-resolution cathodoluminescence (CL) studies reveal both the microstructure of the samples and trace element inclusions. Titanium (Ti) and aluminum (Al) are the two of the most abundant impurities in quartz. Ti is the dominant CL activator while Al is not generally considered an activator \cite{tailby2018,leeman2012study}. Trace element studies show that quartz samples of different geological origins have a wide range of Ti and Al concentrations. Low-temperature hydrothermal vein quartz (HVQ) has the lowest trace element concentrations: quartz typically has a Ti concentration of a few 100 ppm; but this number could be as low as 6 ppm for an HVQ sample, which simultaneously has a low Al concentration \cite{rusk2008trace,ackerson2015trace}. Here, we characterize preliminary measurements to demonstrate that low-Ti vein quartz samples are a suitable choice for our proposed experiment (see Figure \ref{fig:CLimage}).

We note that any mineral deposits with comparable geological history and suitability for low-background CL scanning could in theory serve as a target in a UHDM search.  However, HVQ is well suited for a high sensitivity search: it is available in large quantities with high purity and crystal homogeneity from deposits with well characterized geological history.  Hydrothermal fluid flow is commonly localized along fracture systems, fault systems, and shear zones that can produce vast arrays of quartz veins. When a fracture or fault remains open and under hydrothermal pressure for a sufficient period of time, hydrothermal vein quartz grows as large, euhedral, and high-purity crystals. These properties will enable us to analyze a large net exposure with the proposed protocol, using serial sectioning and SEM-CL scanning of large samples, with background rejection via correlation of damage spots across layers (see Figure \ref{fig:readout}b).


HVQ from the Jack Hills of Western Australia is an ideal source of quartz for the DM search. The siliciclastic units at Jack Hills contain numerous, large quartz veins that appear as prominent surface features (i.e., clusters of milky white outcrops that can form very localized topographic highs) observed throughout the range. The veins are generally either undeformed or very weakly deformed, and often show an abundance of high-purity, gem-quality quartz crystals. These HVQ systems can reach impressive sizes at several locations within the Jack Hills from cm-scale to 50 meters wide \cite{spaggiarietal2007jack}. Several of the vein systems can be followed for several kilometers and appear to be associated with major episodes of brittle faulting. The combined work of Rasmussen et al. \cite{rasmussen2010situ} and Spaggiari \cite{spaggiari2007jack} provide strong evidence that units (including the hydrothermal quartz veins) at Jack Hills, particularly in the vicinity of the "classic" W74 location, have likely remained at temperature conditions less than 330-420 $^{\circ}$C for $\sim$ 1.7 Gyr. The fact that the tectonic environment can be evaluated in detail \cite{trail2016li,cavosie2004internal,baxter1984jack,spaggiari2007jack} (including thermometry and age dating) and consistently demonstrates equilibria at such low temperature (i.e., at or below greenschist facies), is to the best of our knowledge unique to Jack Hills, making it an ideal source of HVQ for our proposed measurements.

\subsection{Experimental Protocol}
\label{sec:protocol}

The proposed experimental protocol is as follows:
\begin{enumerate}
    \item Identify quartz samples that are (i) old, having last annealed no less than 1 Gyr ago; and (ii) clean, with low CL noise level and less than a few thousand micron-scale resolved CL features per cm$^2$.
    \item Prepare about $10^4$ samples of size $\sim \cm^3$ that satisfy the above conditions and prepare each of the samples into sections of thickness $\sim 100~\micron$ (see Figure \ref{fig:readout}a). Polish the top and bottom surfaces of each section, then scan them with SEM-CL.
    \item Search for correlated damage spots, across the first few sections, that are aligned, section-to-section, along a straight line (see Figure \ref{fig:readout}b).
    \item If such a damage track of interest is identified, perform a dedicated search in subsequent sections to reject false positives.
    \item Repeat steps 3 and 4 for all the $\cm^3$ samples.
\end{enumerate}

To prepare HVQ samples for SEM-CL analysis we perform a standard petrographic polishing preparation. Conventional polishing, using successive SiC powders ranging from 240-1200 ANSI\footnote{American National Standards Institute} grit sizes ($\sim$10 minutes per grit) are used to gradually reduce surface topography. SiC polishing is followed by a series of automated polishing stages on a MiniMet polishing/grinding unit, proceeding from 3 to 1 to 0.1 micron Al$_2$O$_3$ grits ($\sim$30-60 mins per grit, with $\sim$1 minute of ultrasonic bathing between grits). As these polishing stages are generally done on automated or semi-automated systems, large volumes of material can be processed. The effects of polishing on the quartz surface is negligible for most aspects of a UHDM search; there may be some localized damage to the near surface environment, restricted to the first few nm \cite{watson2016crystal}, whereas SEM-CL typically probes about $1~\micron$.

Additionally, polishing and SEM scanning should not significantly heat the sample. Paleomagnetic studies on ancient rock samples show that the magnetization, which would change under heating by only a few tens of degrees, remains unchanged after polishing \cite{Weiss791}.  While SEM interrogation may induce some heating for macroscopic, microns-thick samples such as we propose here, such heating should be limited to $\sim 10^\circ{\rm C}$  \cite{holmes2000thermal,wang2019measurement}--especially with appropriately chosen SEM beam parameters. SEM-CL can therefore be considered a non-destructive readout method.

Alternative methods --including high resolution computed tomography x-ray (CT) or superresolution optical microscopy-- offer sufficiently high resolution to measure UHDM-induced damage tracks, as well as the prospect of three-dimensional imaging.  However, these methods generally require much longer scan times than SEM-CL, as well as method-specific constraints such as sample size restrictions or synchrotron beam availability.  If correlated damage spots in serially-sectioned quartz slabs are identified by the proposed SEM-CL readout, follow-up experiments on the same sample could be pursued (due to the non-destructive nature of the SEM technique); e.g., 3D imaging using techniques such as CT or laser scanning confocal microscopy at different wavelengths.

The scanning rate with SEM-CL depends on sample properties such as the concentration of CL activators. Given a typical data acquisition time of $\sim$ 100 min per $\cm^2$ with $\micron$ resolution (for example see Figure \ref{fig:CLimage}), we plan the experiment in three stages:
\begin{itemize}
    \item Quartz-$1 \, \m^2$: About two years of experiment time with four SEM-CL apparatuses will be required to scan samples with a total area of about 1 m$^2$. The total quartz exposure of $\sim 1\, \m^2\,\Gyr$ for such a search would probe UHDMs of mass $m_\DM \lesssim 10^{27}~\GeV$. This first stage search would probe a currently unconstrained mass range with a new technique; see Figure \ref{fig:param_space}.
    \item Quartz-$10 \, \m^2$: 20 SEM apparatuses running for about four years would provide a total quartz exposure to UHDMs of $\sim 10\, \m^2\,\Gyr$, yielding sensitivity $m_\DM \lesssim 10^{28}~\GeV$.
    \item Quartz-$100 \, \m^2$: 100 SEM apparatuses running for about eight years would provide a total quartz exposure to UHDMs of $\sim 100\, \m^2\,\Gyr$, yielding sensitivity $m_\DM \lesssim 10^{29}~\GeV$.
\end{itemize}

\subsection{Model-Independent Sensitivity}
\label{sec:sensitivity}
The proposed 
experiment would be sensitive to a wide range of ultra-heavy dark matter (UHDM) candidates, independent of the underlying dark sector microphysics, that (1) pass through the quartz sample with sufficiently high probability while (2) depositing enough energy in a sufficiently concentrated way to melt a micron-size lateral region. 

Given a DM candidate of mass $m_{\rm DM}$, we can estimate the expected number of DM transits in a sample of area $L\times L$ over a duration $T$ to be
\begin{equation}
    N \sim 1 \left(\frac{10^{29}\text{ GeV}}{m_\DM}\right) \left(\frac{L}{10\,\m}\right)^2 \left(\frac{T}{10^9\text{ year}}\right)\label{eqn:eventrate}
\end{equation}
based on the local DM density, $\rho_{\rm DM}\approx 0.3\,\GeV/\text{cm}^3$. As described in the previous sections, the quartz samples under consideration are roughly $T\sim 10^{9}\text{ year}$ old, and a 100 $\m^2$ sample area can be scanned in stage three. The requirement that $N\gtrsim 1$ imposes an upper bound on the UHDM mass, $m_{\rm DM}\lesssim 10^{29}\text{ GeV}\sim 100\text{ kg}$. The advantage afforded by the large exposure of such a long-lived sample is manifest. 

An UHDM moving through the Earth will collide with and deposit energy to SM particles along its path. The energy $E_1$ imparted to each SM nucleus can go as high as the kinematical limit of $10~\keV$ (corresponding to nuclei acquiring twice the velocity of the UHDM in a collision) depending on how elastic these collisions are, while the stopping power $dE/dx$ depends on $E_1$ as well as the UHDM radius. For simplicity, we assume in our estimates that the UHDM travels at least a few kilometers deep into the Earth's surface while maintaining its Milky Way virial velocity of $v_{\rm DM}\sim 10^{-3}c$. This amounts to an upper bound on the energy deposition rate
\begin{equation}
    \frac{dE}{dx}\lesssim 10^{13}\frac{\MeV}{\angs}\left(\frac{m_\DM}{10^{29}\text{ GeV}}\right).\label{eqn:dragforcebound}
\end{equation}
Most of the deposited energy will likely go to SM nuclei. Only a tiny portion will go directly to electrons, whose low mass limits their kinetic energy gain (for kinematics reasons) and whose coupling to DM is severely limited by astrophysical and cosmological constraints \cite{Green:2017ybv}. The nuclei and electrons will then thermalize, leading to a loosening of molecular bonds as the electrons acquire more energy, and eventually cause melting. Due to thermal diffusion, the melted region will enlarge and cool. What ultimately remains, in the case of quartz, is a long cylindrical trail of amorphous silica, precisely the kind of damage that is detectable with the method outlined above.

\begin{figure}[htbp!]
    \centering
    \includegraphics[width=\columnwidth]{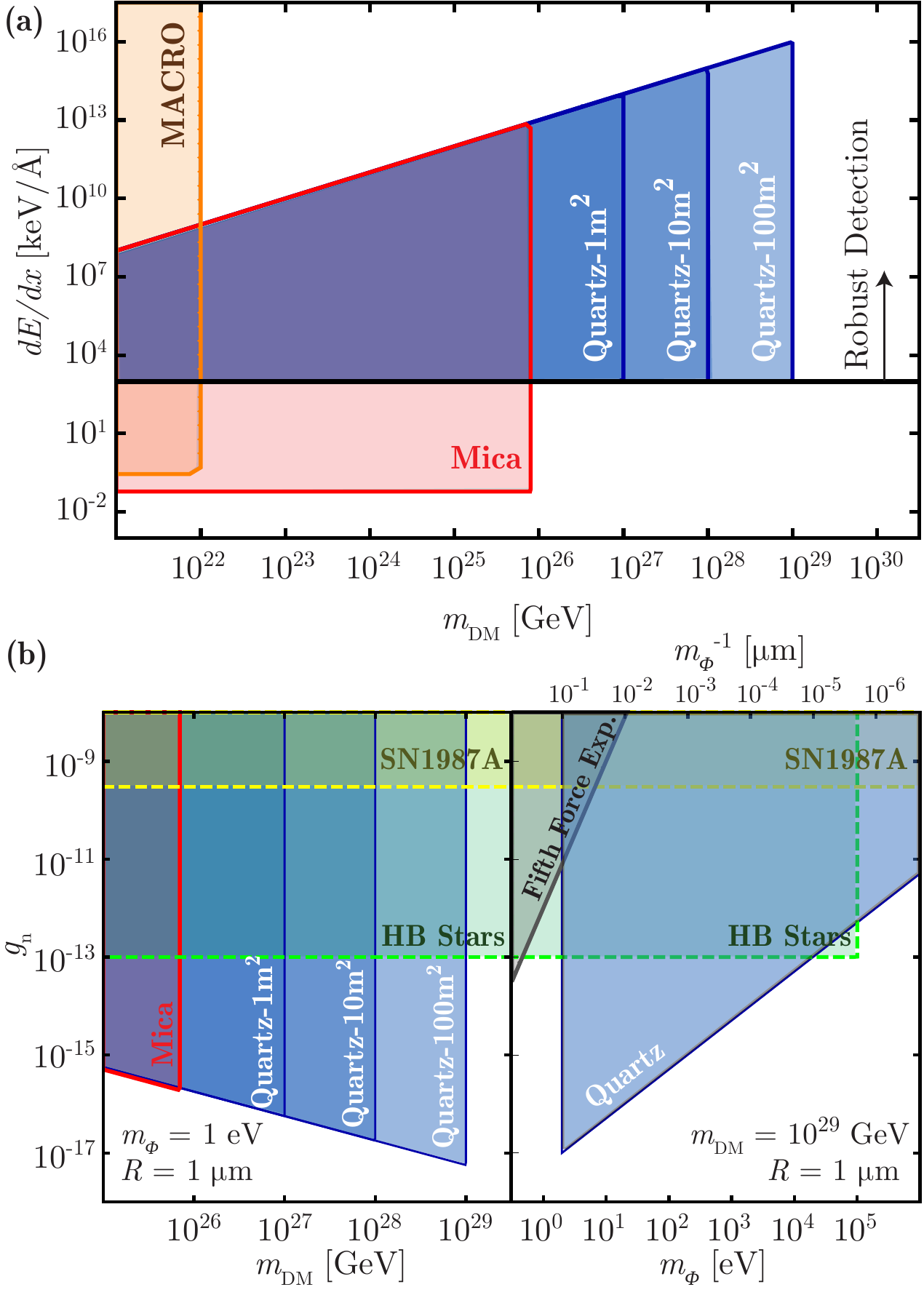}
    \caption{{Sensitivity projections for the proposed ultra heavy dark matter (UHDM) search.} {\bf (a)} Model-independent reach of the geological-quartz detector proposal expressed as stopping power $dE/dx$ vs mass $m_{\rm DM}$ of a passing UHDM particle, together with the existing constraints from MACRO for energy deposition per nucleus $E_1\sim 1~\eV$  \cite{Ambrosio:2004ub, Scholz:2016} as well as from damage track searches in ancient mica \cite{price1986}. The vertical and slanted boundaries of the quartz-detectable parameter space (for different effective detector areas) stem from the requirements of an $O(1)$ probability of transit, Eq.~\eqref{eqn:eventrate}, and a negligible slowing of the UHDM up to a $1\text{ km}$ depth, Eq.~\eqref{eqn:dragforcebound}, respectively. The black horizontal line indicates the melting threshold for a micron-sized lateral region, Eq.~\eqref{eqn:meltingenergydeposition}, above which robust detection is possible. {\bf (b)} Parameter space of the UHDM model considered in Sec.~\ref{sec:model}. \textit{Left:} reach on coupling $g_{\rm n}$ vs DM mass $m_{\rm DM}$. \textit{Right:} reach on coupling $g_{\rm n}$ vs mediator mass $m_\phi$. Also shown are existing constraints from ancient mica \cite{price1986}, fifth force experiments \cite{Green:2017ybv}, and stellar cooling of SN1987A \cite{Green:2017ybv} and horizontal branch (HB) stars \cite{Green:2017ybv} (note that the stellar cooling bounds are model-dependent \cite{DeRocco:2020xdt}). In these $g_{\rm n}$ plots, we set $g_\chi$ to its upper bound $m_\phi/\Lambda_\chi$ from Eq.~\eqref{stability}.}
    \label{fig:param_space}
\end{figure}

In order to leave a robustly detectable damage trail, the UHDM must deposit sufficient energy per unit length $dE/dx$ exceeding the required latent heat to melt each unit-length segment of a micron-radius cylinder. This amounts to a $dE/dx$ threshold for robust detection of
\begin{equation}
    \frac{dE}{dx}\gtrsim \frac{\MeV}{\angs}\,.\label{eqn:meltingenergydeposition}
\end{equation}
See Figure \ref{fig:param_space}a for model-independent sensitivity projections. Further, since quartz has a melting point of $10^4~\rm K \sim 1~\eV$ and energy tends to spread outward, the energy deposition must be sufficiently localized that the energy $E_1$ gained by each nucleus is greater than the melting temperature, i.e. it must lie in the range
\begin{equation}
   1\,\eV\lesssim E_1 \lesssim 10\, \keV \label{eqn:meltingenergypernucleus}
\end{equation}
where the upper bound is the kinematical limit for energy transfer per nucleus (with mass number $A\sim 10$).

\section{Example UHDM Model}
\label{sec:model}

In this section, we consider an example of a simple, ultra-heavy composite dark matter (UHDM)  state \cite{Grabowska:2018lnd} that can give rise to the desired damage tracks while being consistent with existing experimental and observational constraints. These composite objects consist of $N_\chi$ dark fermions $\chi$ whose mass, inverse size, and binding energy to form the UHDM are determined by a single scale $\Lambda_\chi$. It follows that they have mass $m_\DM \sim N_\chi \Lambda_\chi$ and size $R \sim N_\chi^{1/3} \Lambda_\chi^{-1}$. We assume that the fermions $\chi$ interact with standard model nucleons $\psi_{\rm n}$ through a repulsive\footnote{Attractive DM-nucleon interactions are just as compelling as the repulsive interaction considered here. We note that the attractive interactions might have more complicated dynamics as nuclei may get trapped and accumulate inside the UHDM. See e.g. \cite{Acevedo:2020avd}}. Yukawa interaction mediated by a scalar $\phi$ of mass $m_\phi$:
\begin{equation}
    -\mathcal{L}\supset \frac{1}{2}m_\phi^2\phi^2+g_{\rm n}\phi\bar{\psi}_{\rm n}\psi_{\rm n}-g_{\chi}\phi\bar{\chi}\chi\,.
\end{equation}
We show that UHDMs with the following properties satisfy the robust detectability criteria detailed in Sec.~\ref{sec:sensitivity} without running afoul of any existing constraints:\footnote{Due to various constraints, this parameter space has a complicated geometry. Here we simply identified the lower and upper limits for each parameter.}
\begin{align}
    10^{26} \,\GeV \lesssim& m_\DM \lesssim 10^{29} \,\GeV\label{eqn:massrange}\\
    10 \text{ nm}\lesssim& R\lesssim 1 \text{ cm}\label{eqn:radiusrange}\\
    0.1\,\eV\lesssim& m_\phi\lesssim \MeV \,\leftrightarrow\, \,
    10^2\text{ fm}\lesssim m_\phi^{-1}\lesssim 1\,\micron
    \label{eqn:mediatorrange}\\    
    100 \,\keV \lesssim& \Lambda_\chi \lesssim 10 \,\GeV\,.\label{eqn:Lambdarange}
\end{align}
This allows us to probe wide ranges of the couplings $g_{\rm n}$ and $g_{\chi}$. Two slices of this parameter space are shown in Figure~\ref{fig:param_space}b. Eq.~\eqref{eqn:massrange} follows from Eq.~\eqref{eqn:eventrate} and ancient mica constraints; Eq.~\eqref{eqn:radiusrange} follows from 
\eqref{eqn:meltingenergydeposition}, \eqref{eqn:meltingenergypernucleus}, \eqref{eqn:E1dEdx}, and the quartz sample size of 1 cm; Eq.~\eqref{eqn:mediatorrange} follows from fifth force constraints and the requirement that the UHDM-nucleus interaction be treated classically; Eq.~\eqref{eqn:Lambdarange} follows from Eqs.~\eqref{eqn:massrange} and \eqref{eqn:radiusrange}.

\subsection{Detectability with Quartz}
The optimal UHDM detection signature is expected for mediators with a range $m_\phi^{-1}$ satisfying $\Lambda_\chi^{-1}\ll m_\phi^{-1} \lesssim \micron$, since this is the intermediate regime where the UHDM-nucleon coupling is enhanced by the number of constituents $\chi$ of the UHDM within the range of the mediator $(m_\phi^{-1}/\Lambda_\chi^{-1})^3$ while simultaneously evading existing fifth force constraints. For simplicity of analysis we only consider part of the parameter space where $m_\phi^{-1}\ll R$. In doing so, we limit the UHDM's cross section to be at most geometrical.

An SM nucleus located inside the UHDM only sees the composite DM particle's constituents $\chi$ within the range of the mediator $ m_\phi^{-1}\ll R$. Hence, to the SM nucleus each point in the bulk of the UHDM is just like any other, yielding a potential energy $V(r)$ as a function of the distance $r$ from the center of the UHDM with the following profile:
\begin{equation}
    V(r)=\begin{cases}
    +V_0, &r<R\\
    0, &r>R
    \end{cases}
\end{equation}
where at the boundary $r\approx R$ the potential drops to zero exponentially over a length scale of order $m_\phi^{-1}$, and 
\begin{align}
    V_0\sim  \left(\frac{\Lambda_\chi}{m_\phi}\right)^3\frac{g_{\chi}(10g_{\rm n})}{m_\phi^{-1}} \label{eqn:V0}
\end{align}
for SM nuclei with mass number  $A\sim 10$. As a result, from the perspective of a nucleus the UHDM is just a constant potential hill moving at a velocity $v_{\rm DM}\sim 10^{-3}c$.

Since the de Broglie wavelengths $(10\,\MeV)^{-1}$ of the SM nuclei are smaller than the mediator ranges $m_\phi^{-1}$ of interest, we can treat the UHDM-nucleus interactions classically. When $V_0\gtrsim 10 \,\keV$, the potential $V_0$ prevents any nucleus from entering the UHDM. The UHDM-nucleus collisions are thus elastic, and the energy $E_1$ transferred to a nucleus saturates the kinematical limit $E_1\sim 10\, \keV$. If $V_0\lesssim 10 \,\keV$, on the other hand, the nuclei can easily climb the potential hill, and the collisions between a nucleus and the UHDM's surface will be inelastic. When a nucleus encounters the surface of the UHDM, it receives a force $F\sim V_0/m_\phi^{-1}$ due to the gradient of the Yukawa potential. This force is exerted throughout the duration $\tau\sim m_\phi^{-1}/v_{\rm DM}$ of the collision, resulting in a nearly-instantaneous momentum kick $p_1\sim F\tau$ which translates to the kinetic energy $E_1\sim 10\,\keV\left(V_0/10\,\keV\right)^2$ per nucleus. To sum up, the energy imparted to a nucleus after the passage of a UHDM is
\begin{equation}
    E_1\sim 10\,\keV \times \text{min}\left[1,\left(\frac{V_0}{10 \,\keV}\right)^2\right].\label{eqn:E1V0}
\end{equation}
Using a lattice spacing of about $5\angs$ for quartz, the energy deposition rate then follows:
\begin{equation}
    \frac{dE}{dx}\sim \frac{E_1}{5\angs}\left(\frac{R}{5\angs}\right)^2. \label{eqn:E1dEdx}
\end{equation}
Having linked the model parameters with the quantities characterizing quartz damage tracks, the detectable parameter space can be evaluated based on the considerations in Sec.~\ref{sec:sensitivity} (see Figure \ref{fig:param_space}b).

\subsection{Existing Constraints}
\label{sec:existing_constraints}
\subsubsection{The mediator}
Past experiments and observations have placed limits on the coupling $g_{\rm n}$ of the mediator $\phi$ to standard model nucleons with varying severity for different masses $m_\phi$ of the mediator. These include: collider constraints on the meson decay rate, laboratory {\it fifth-force} searches, and stellar cooling bounds from observations of the SN1987A event and horizontal branch (HB) stars. Note, however, that the stellar cooling bounds are model-dependent \cite{DeRocco:2020xdt}. The following parameter space is thus ruled out \cite{Green:2017ybv}:

\begin{itemize}
    \item Meson decay: $g_{\rm n}\gtrsim 10^{-6}$, $m_\phi\lesssim 100\,\MeV$.
    \item Fifth force: $g_{\rm n}\gtrsim 10^{-12} (m_\phi/\eV)^3$, $m_\phi\lesssim 100\,\eV$.
    \item SN1987A: $3\times10^{-10}\lesssim g_{\rm n}\lesssim 3\times10^{-7}$, $m_\phi\lesssim 30\,\MeV$.
    \item HB stars: $g_{\rm n}\gtrsim 10^{-13}$, $m_\phi\lesssim 100\,\keV$.
\end{itemize}
Furthermore, the couplings of UHDM constituents $\chi$ to the mediator $\phi$ add extra self-interactions among $\chi$ that may destabilize the UHDM. In order for the UHDM to be stable the mediated self-interaction potential $g_\chi^2\Lambda_\chi^3m_\phi^{-2}$ of a single $\chi$ must not exceed the binding energy $\Lambda_\chi$. This puts an upper bound on the coupling $g_\chi$ of $\chi$ to the mediator $\phi$:
\begin{equation}
    g_{\chi}\lesssim \frac{m_\phi}{\Lambda_\chi}.\label{stability}
\end{equation}

\subsubsection{Direct detection}
Of the currently and previously running direct-detection DM experiments, MACRO puts a strong constraint on our scenario due to its large volume. MACRO is a scintillator experiment with exposure of about $10^3\text{ m}^2\times 10\text{ years}$ corresponding to $m_\DM \lesssim 10^{22}\,\GeV$ for 1 event over its decade-long lifespan. It is sensitive to energy depositions $\gtrsim 10\,\MeV/\cm$ \textit{to electrons} \cite{Ambrosio:2004ub}. When a nucleus receives energy $E_1$ from interaction with a UHDM, only some fraction $Q(E_1)$, called the quenching factor, of that energy effectively goes to the electrons tied to the nucleus. It is this relatively small fraction of energy that is responsible for the processes of scintillation and ionization that may occur subsequently. We can translate MACRO's $10\,\MeV/\cm$ detection threshold to a sensitivity \textit{to nuclear energy depositions} via effecting an increase by the quenching factor $Q(E_1)$ \cite{Scholz:2016}. 

An even more stringent bound on our model arises from direct searches for long damage tracks in muscovite mica crystals \cite{price1986}. The non-observation of tracks extending beyond naturally occurring defects and radioactivity damage was originally used to constrain the abundance of magnetic monopoles, but also limits the UHDM parameter space. This past mica search involved total sample area $\sim 1200 ~\cm^2$ with sample ages $\simeq 5 \times 10^8 \,\yr$, corresponding to a DM reach of $m_\DM \lesssim 10^{26}~\GeV$. The energy deposition threshold for detection in this experiment via etching and optical microscopy was identified as $dE/dx \gtrsim 6~\GeV/\cm$.    

\subsubsection{Astrophysical and Cosmological limits}
Indirect limits can also be placed on the couplings $g_{\rm n}$ and $g_{\chi}$ from the limits on DM-baryon and DM-DM cross sections. DM-baryon interactions in the early universe can affect baryon acoustic oscillations and is therefore constrained by CMB and LSS observations. This puts an upper bound on the DM-baryon momentum-transfer cross section that would be observed today: $\sigma_{\chi\rm b}/m_{\rm DM}\lesssim 10^{-3}\text{ cm}^2/\text{g}$ \cite{Dvorkin:2013cea}. Astronomical observations of the Bullet Cluster also place a limit on the DM-DM momentum-transfer cross section: $\sigma_{\chi\chi}/m_{\rm DM}\lesssim 1\text{ cm}^2/\text{g}$ \cite{Spergel:1999mh}. Since we are mainly interested in UHDMs with geometrical cross sections of order $\micron^2$ and masses up to $10^{29}~\GeV (100 ~{\rm kg})$, these astrophysical and cosmological observations only impose significant constraints on the low mass side of our parameter space. Moreover, these constraints are alleviated if UHDMs constitute less than $10\%$ of the total DM mass, in which case the maximum detectable mass would also be lowered by an order of magnitude.

\section{Conclusion and Outlook}
\label{sec:conclusion}

Given the diverse range of theoretically well-motivated dark sectors, it is critical to perform searches with techniques that are sensitive to a broad class of dark-sector phenomena. In this paper, we propose a detection method for ultra-heavy composite dark matter particles (UHDMs). Our proposed experiment is based on mapping damage tracks in ancient quartz samples with SEM-CL scanning. This method has two significant advantages: (1) the billion-year exposure time of such samples enables us to probe DM candidates with masses as high as $10^{29}~\GeV (100 ~{\rm kg})$, surpassing the reach of existing direct-detection experiments, and (2) the distinctly-long cylindrical damage trails left by such UHDMs are easily distinguished from other features at the relevant scales.

In this work, we focus on detecting long tracks of amorphous silica in quartz samples expected from passing UHDMs that impart enough energy to cause melting. In future work, we will consider the feasibility of extending the experimental sensitivity to energy deposition rates below the melting threshold. For that purpose, we intend to carry out a number of studies including: (i) signal calibration by artificially creating damage tracks in synthetic quartz samples with a high-power pulsed laser of variable intensity and comparing it with the resulting CL signal levels; and (ii) noise calibration by preparing a set of quartz samples, natural and synthetic, with different concentrations of CL activators and analysing their CL emission rates. These studies will provide us a better understanding of the signal-to-noise ratio as seen in SEM-CL imaging, which will allow us to better estimate the detection threshold.

We focus here on  quartz as a promising target mineral for UHDM detection because of the availability of abundant high-quality, old samples and the suitability of SEM-CL as a high-resolution, fast readout method. However, any mineral satisfying these two criteria is a potential target material; identifying such targets is valuable in verifying putative UHDM signals in a complementary search,  possessing different systematics.

Our proposed experiment is largely agnostic to the detailed microphysics of the dark sector, as long as it results in long damage tracks in geological quartz. To demonstrate the projected reach of the proposed approach, we considered a QCD-like dark sector that interacts with the standard model repulsively via a light mediator. The particle spectrum of this theory includes heavy bound states, composed of a large number of elementary dark fermions, which could create interesting targets for detection. We identified experimentally-detectable regions of the parameter space that satisfy various limits derived from phenomenological considerations as well as past observations. In future work, it would be interesting to delineate a broader range of DM models than can lead to similar damage patterns in ancient rock.


\begin{acknowledgments}
This work was supported by the DOE QuANTISED program under Award No. DE-SC0019396; the Army Research Laboratory MAQP program under Contract No. W911NF-19-2-0181; and the University of Maryland Quantum Technology Center. SR is supported by the NSF under grant PHY-1818899, the SQMS Quantum Center and DOE support for MAGIS. 
\end{acknowledgments}

\bibliographystyle{apsrev4-1}
\bibliography{references}

\end{document}